\documentclass[
 reprint,
 amsmath,
 amssymb,
 aps,
]{revtex4-1}

\usepackage[pdftex]{graphicx}
\usepackage{wasysym}
\usepackage{amsfonts}
\usepackage{ifsym}

\newcommand{\nn}{\nonumber}

\newcommand{\COMMENT}[1]{}

\newcommand{\neqa}{\nonumber\end{eqnarray}}
\newcommand{\la}[1]{\label{#1}}

\newcommand{\<}{{\langle}}
\renewcommand{\>}{{\rangle}}

\newcommand{\re}{\relax{\rm I\kern-.18em R}}

\def\su2{{SU(2)}}

\def\[{\left[}
\def\]{\right]}

\def\({\left(}
\def\){\right)}
\def\[{\left[}
\def\]{\right]}

\def\<{\langle}
\def\>{\rangle}

\def\i2{\frac{i}{2}}

\def\cP{{\cal P}}

\def\2F1{\,_2{\rm F}_1}

\newcommand{\ft}[2]{{\textstyle\frac{#1}{#2}}}

\usepackage[usenames,dvipsnames]{xcolor}
\usepackage{color}
\usepackage{graphicx}
\usepackage{dcolumn}
\usepackage{bm}
\usepackage{hyperref}

\newcommand{\beq}{\begin{equation}}
\newcommand{\eeq}{\end{equation}}
\newcommand{\beqq}{\begin{equation*}}
\newcommand{\eeqq}{\end{equation*}}
\newcommand\beqa{\begin{eqnarray}}
\newcommand\eeqa{\end{eqnarray}}
\newcommand\beqaa{\begin{eqnarray*}}
\newcommand\eeqaa{\end{eqnarray*}}
\newcommand\bea{\begin{array}}
\newcommand\eea{\end{array}}

\begin{document}

\title{Space-time S-matrix and Flux-tube S-matrix at Finite Coupling}

\author{Benjamin Basso$^{\displaystyle\pentagon}$, Amit Sever$^{{\displaystyle\pentagon},{\displaystyle\Box}}$ and Pedro Vieira$^{\displaystyle\pentagon}$}

\vspace{15mm}

\affiliation{$^{\displaystyle\pentagon}$Perimeter Institute for Theoretical Physics,
Waterloo, Ontario N2L 2Y5, Canada\\
$^{\displaystyle\Box}$School of Natural Sciences, Institute for Advanced Study, Princeton, NJ 08540, USA
}

\begin{abstract}
We propose a non-perturbative formulation of planar scattering amplitudes in ${\cal N}=4$ SYM or, equivalently, polygonal Wilson loops. 
 The construction is based on the OPE approach and introduces a new decomposition of the Wilson loop in terms of fundamental building blocks named {\it Pentagon transitions}. 
These transitions satisfy a simple relation to the worldsheet S-matrix on top of the so called Gubser-Klebanov-Polyakov vacuum which allows us to bootstrap them
at any value of the coupling. In this letter we present a subsector of the full solution to scattering amplitudes which we call {\it the gluonic part}. We match our results with both weak and strong coupling data available in the literature. For example, the strong coupling  Y-system can be understood in this approach. 

\end{abstract}

\maketitle

\section{Introduction} 
Computing the full S-matrix of a four dimensional gauge theory at finite coupling might seem impossible. Conventional techniques, based on perturbation theory, soon become too cumbersome as the number of loops increases. Besides, the final results are typically much simpler than the intermediate steps would suggest. Both observations beg for an alternative non-perturbative approach. In the large $N_c$ expansion, a dual \textit{two dimensional} string theory of  't Hooft surfaces emerges as such an alternative description. In some cases, these 't Hooft surfaces are integrable and their dynamics can be studied exactly. This is what happens in $\mathcal{N}=4$ SYM theory and has led to the full solution of the problem of computing all two point correlation functions of local operators \cite{review}. 
Higher point correlation functions, Wilson loops (WL) and scattering amplitudes are considerably richer objects that depend on several external kinematics and probe string interactions. Since the string material is the same we expect integrability to help us compute these observables at any value of the coupling as well. 

In this paper we consider planar Scattering Amplitudes or Null Polygon WLs in $\mathcal{N}=4$ SYM (in this theory they are the same \cite{AM,AmplitudeWilson,superloop}).  We identify a new object, called \textit{Pentagon transition}, as the building block of these WLs. The Pentagon transitions arise naturally in the OPE construction \cite{OPEpaper} and completely determine the WL at any coupling. Remarkably, these transitions are directly related to the dynamics of the Gubser-Klebanov-Polyakov (GKP) flux tube \cite{GKP,AldayMaldacena} and can be computed exactly using Integrability! In this paper we present the most fundamental ones, describing the transition of gluonic degrees of freedom. 

\section{Framing the Wilson loop}

Our construction is based on a decomposition of a general polygon WL into simpler fundamental building blocks which we will denote as \textit{square} and \textit{pentagon} transitions.

\begin{figure}[h]
\centering
\def\svgwidth{8.5cm}
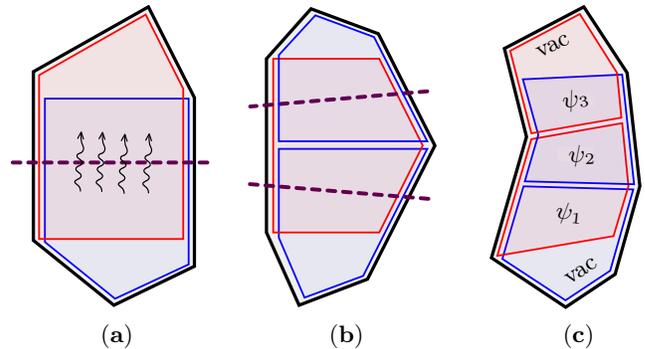
\caption{Decomposition of $n$-sided Null Polygons into sequences of $n-3$ null squares. Any two adjacent squares form a pentagon and any middle square is shared by two pentagons. There are $n-4$ pentagons and $n-5$ middle squares. Every middle square in the decomposition shares two of its opposite cusps with the big polygon; the positions of the other two cusps (which are not cusps of the big polygon) are fixed by the condition that they are null separated from their neighbours. For example, in (a) we have an hexagon. It has a single middle square whose symmetries $\tau,\sigma$ and $\phi$ parametrize its three conformal cross-ratios \cite{OPEpaper}.  
 }\label{HexagonHeptagonOctagon}
\end{figure}

We decompose a polygon into a sequence of null squares as in figure \ref{HexagonHeptagonOctagon}. Any two adjacent squares form a \textit{pentagon}. 

\begin{figure}[t]
\centering
\def\svgwidth{8.5cm}
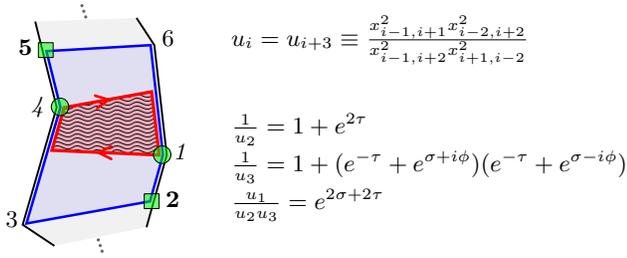
\caption{For any middle square in the framing we associate a GKP time $\tau$, space $\sigma$, and angle $\phi$ for rotations in the two dimensional space transverse to it. They are the three conformal cross ratios associated with an hexagon that is formed by the two pentagons overlapping on that square. Note that cusps $\bf 2$ and $\bf 5$ are the only cusps of the hexagon that are not shared with the big polygon. Cusps $\it 1$ and $\it 4$ are the only cusps of the hexagon that are shared with the middle square. 
An hexagon is symmetric under $\phi \leftrightarrow -\phi $. The relative sign between $\phi_i$ and $\phi_{i+1}$ is physical and fixed by demanding that in the measure limit, $\sigma_i,\sigma_{i+1}\to-\infty$, they only appear in the combination $\phi_i+\phi_{i+1}$.
}\label{car}
\end{figure}

Of particular importance are the \textit{middle squares} that arise as overlap of two consecutive pentagons. 
For an $n$-edged polygon there are $n-5$ middle squares. Each of them has three symmetries parametrized by a GKP time $\tau_i$, space $\sigma_i$, and angle $\phi_i$ for rotations in the two dimensional space transverse to this middle square. We coordinatize all conformally inequivalent polygons by acting with the symmetries of the $i$-th middle square on all cusps to the bottom of that square~\cite{heptagonPaper}. The set $\{\tau_i,\sigma_i,\phi_i\}_{i=1}^{n-5}$ parametrizes the $3n-15$ independent conformal cross ratios of a $n$-edge null polygon. An explicit definition is given in figure~\ref{car}. 

\begin{figure}[t]
\centering
\def\svgwidth{8cm}
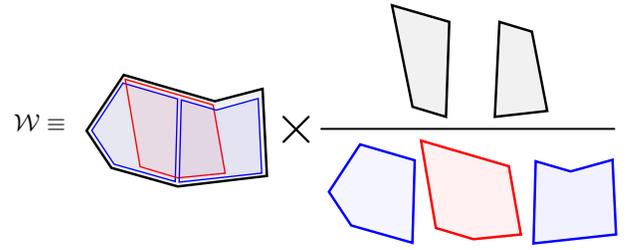
\caption{We construct a conformal invariant finite ratio by dividing the expectation value of the WL by all the pentagons in the decomposition and multiplying it by all the middle squares, $\mathcal{W}\equiv \<W\> \times \frac{\<W_{1^\text{st} \text{middle sq.}}\>\<W_{2^\text{nd} \text{middle sq.}}\>\dots }{ \< W_{1^\text{st} \text{pent.}}\>  \< W_{2^\text{nd} \text{pent.}}\>   \dots } $. This is a generalization of the ratios considered in \cite{OPEpaper,heptagonPaper}.  
 }\label{calW}
\end{figure}

We regulate the well understood UV divergences of the WL using pentagons and squares as defined in figure \ref{calW}. These squares and pentagons have no conformal cross ratios; their expectation values are fixed by conformal symmetry \cite{anomaly} and given by the BDS ansatz \cite{BDS}. Therefore, we lose no information by considering these conformal invariant finite ratios $\cal W$.

Having described the kinematics we now move to the dynamics, depicted in figure \ref{HexagonHeptagonOctagon}c. 
We start with the GKP vacuum in the bottom and evolve it all the way to the top where it is overlapped with the vacuum again. In between,  we decompose the flux tube state in the $i$-th middle square over a basis of GKP eigenstates $\psi_i$. 
Each eigenstate $\psi_i$ propagates trivially in the corresponding square for a time $\tau_i$. It then undergoes a Pentagon transition $\mathcal{P}$ to
the consecutive square 
where it is decomposed again
and so on: 
\beq
\text{vacuum}\,  \to\,\psi_1\, \to \dots \to\, \psi_{n-5}\,\to\, \text{vacuum} \, .
\eeq 
In particular for the hexagon we have vacuum $\to \psi_1 \to $ vacuum while for the heptagon we have vacuum $\to \psi_1 \to \psi_2 \to$ vacuum. 
We see that the heptagon is the first polygon that contains non-trivial transitions between arbitrary states; for the hexagon the transitions always involve the vacuum.

Following this picture we can write any $n$-sided WL as 
\beqa\la{decomposition}
\mathcal{W}=&&\sum_{\psi_i} e^{\sum_{j}\(-E_j\tau_j+ip_j\sigma_j+im_j\phi_j\)}\times  \\ &&\cP(0|\psi_1) \cP({\psi}_1|\psi_2)\dots \cP({\psi}_{n-6}|\psi_{n-5})\cP({\psi}_{n-5}|0) \,.\,\,\, \nn
\eeqa
The eigenstates $\psi_i$ have definite energy $E_i$, $U(1)$ charge $m_i$, and momenta $p_i$. They are $N$-particle states with $N$ excitations on top of the GKP flux tube with $N=0,1,2,\dots$. The charges $E_i,m_i,p_i$ of the eigenstate are the sum of the charges of the individual excitations. A useful way to parametrize the energy and momentum of any excitation is through a Bethe rapidity $u$. Then each state is parametrized by a set of rapidities ${\bf u}=\{u_1,\dots,u_N\}$. Furthermore, the GKP excitations can be fermions, gluons, scalars or bound-states of different excitations \cite{BenDispPaper}; we use $a_j$ to indicate which kind of excitation the $j$-th particle is and ${\bf a}=\{a_1,\dots,a_N\}$ to parametrize the state. 
We can now be even more explicit and re-write (\ref{decomposition}) using these labels. We shall do it for the hexagon and heptagon since the generalization is obvious. 
We have 
\beqa
\mathcal{W}_\text{hex}&=&\sum_{{\bf a}} \int\!\! d{\bf u} \,P_{\,\bf a}(0|{\bf u})  P_{\,{\bf a}}(\bar{\bf u}|0) \, e^{-E({\bf u})\tau+ip({\bf u})\sigma+im\phi}  \,,\,\,\,\,\nonumber \\
\mathcal{W}_\text{hep}&=&\sum_{{\bf a},{\bf b}} \int \!\! d{\bf u}\,d{\bf v} \,P_{\,\bf a}(0|{\bf u}) P_{\,{\bf a}{\bf b}}(\bar{\bf u}|{\bf v}) P_{\,{\bf b}}(\bar{\bf v}|0) \la{heptagondecomposition} \label{Whexhept}\\ && e^{-E({\bf u})\tau_1+ip({\bf u})\sigma_1+im_1\phi_1-E({\bf v})\tau_2+ip({\bf v})\sigma_2+im_2\phi_2}  \,\,\,\,\,\nonumber
\eeqa
where $\bar {\bf u}=\{-u_N,\dots,-u_1\}$ and the measure is
\beq
d {\bf u} ={\cal N}_{\bf a} \prod_{j=1}^N \mu_{a_j} (u_j) \frac{du_j}{2\pi} 
\eeq
 and similar for $d {\bf v}$. Here, ${\cal N}_{\bf a}$ is a symmetry factor. It is equal to $1/N!$ for identical particles for example.
 
\begin{figure}[t]
\centering
\def\svgwidth{6cm}
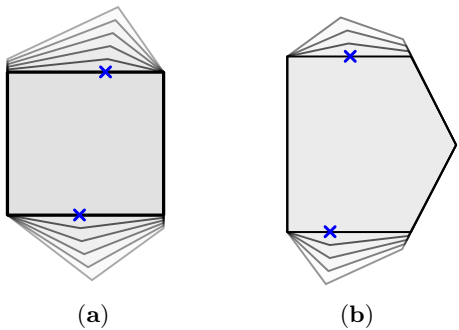
\caption{Two fundamental building blocks: the expectation value of the square ({\bf a}) and pentagon WL ({\bf b}) with GKP excitations inserted on their bottom and top. One natural way to insert these excitations is to start from the hexagon or heptagon (regulated as in fig.\ref{calW}) and take the collinear limit $\tau_i\to\infty$. In this way we can extract the transitions from known Amplitudes/WL in perturbation theory and match them with the integrability predictions; more details in \cite{toappear}.  
}\label{CollinearLimit}
\end{figure}

 The measure and the pentagon transitions are not independent. Instead they are related as
 \beq\la{measureEq}
\underset{v=u}{\operatorname{{\rm Res}}}\, P_{aa}(u|v)= \frac{i}{\mu_a(u)}   \,. 
 \eeq
This relation is understood as follows. We can think of the Pentagon transitions as pentagon Wilson loops with insertions, see figure \ref{CollinearLimit}b. In position space, taking the residue $u=v$ is equivalent to studying the $\sigma_1,\sigma_2\to-\infty$ limit of the pentagon transition with $\sigma_1-\sigma_2$ fixed. This limit corresponds to sending the bottom and top pentagon insertions to the edge opposite to the middle cusp, i.e. close to the left edge in figure \ref{CollinearLimit}b. This is conformally equivalent to flattening the right cusp in this figure. In this way we end up with the square depicted in figure \ref{CollinearLimit}a. This relates the heptagon and hexagon expansions and translates into (\ref{measureEq}).

Contrary to bare pentagon WL (with no insertions), the Pentagon transitions are no longer fixed by conformal symmetry. Remarkably enough, as we will see below, they can be fixed exactly using Integrability.

\section{The Pentagon Transition}
\begin{figure}[h]
\centering
\def\svgwidth{8.5cm}
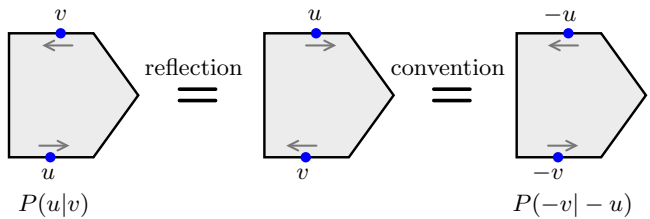
\caption{Flipping the sign of both momenta is equivalent to a reflection of the pentagon. }\label{Reflection2}
\end{figure}

We work with the normalization where the creation amplitude for a single particle is set to one $P_a(0|u)=1$. We start by considering gluonic transitions involving the twist one excitations $F\equiv F_{z-}$ and $\bar F \equiv F_{\bar z-}$, even though most formulae hold untouched for any kind of excitation as discussed later. We denote $P(u|v)\equiv P_{FF}(u|v)$ and $\bar P(u|v)\equiv P_{F\bar F}(u|v)$.
We now postulate three main axioms that single particle transitions should obey. 
The first axiom follows from the reflection symmetry of the pentagon as depicted in figure \ref{Reflection2}.
It reads
\begin{equation}\label{parity}
P(-u|-v) = P(v| u) \,. 
\end{equation}
\vspace{-.7 cm}
\vspace{-.3 cm}

We dub the second axiom as the \textit{fundamental relation}:
\begin{equation}\label{fundamental}
P(u|v) = S(u, v)P(v| u) \,. 
\end{equation}
where $S(u,v)$ is the GKP S-matrix for the scattering of two $F$ excitations. 
Similarly $\bar P(u|v) = \bar S(u, v)\bar P(v| u)$ where $\bar S$ is the scattering phase between an $F$ and an $\bar F$ excitation. (It turns out that the two gluonic S-matrices are related as $(u-v-i)S(u,v)=(u-v+i)\bar S(u,v)$.) All S-matrices, between any pair of excitations, can be computed exactly using integrability \cite{toappear,toappearAdam}, following \cite{BenDispPaper}. The fundamental relation (\ref{fundamental}) establishes a precise bridge between the worldsheet S-matrix $S(u,v)$ and the space-time S-matrix which is built out of pentagon transitions. 

\begin{figure}[h]
\centering
\def\svgwidth{8.5cm}
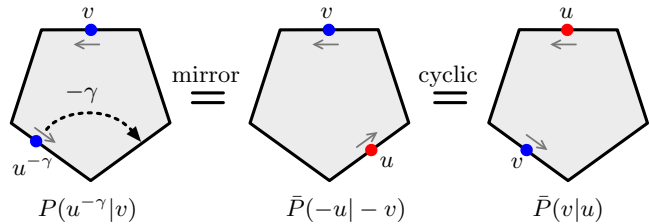
\caption{Under a mirror transformation $u\to u^{-\gamma}$ an excitation is sent to the neighboring edge on its right. This is consistent with exchanging GKP space and time (in the bottom square). Combining this transformation with a cyclic rotation leads to (\ref{crossingEq}).
Under the mirror transformation the gluonic transition $P$ becomes a $\bar P$ transition. This is just kinematics: after mirror we have a new decomposition of the pentagon into two squares. In new tessellation, the relative $U(1)$ charge of the excitation is flipped. This justifies some occurrences of $\bar P$ versus $P$ in the main text, see e.g. (\ref{crossingEq}).}\label{mirror}
\end{figure}

The reader familiar with Watson equations for form factors \cite{Watson} will be tempted to draw an analogy between the fundamental relation (\ref{fundamental}) and similar relations arising in that context. This analogy is however a bit dangerous since in our case one excitation is in the bottom while the other is in the top of the pentagon. If both were in the bottom (or in the top) then it would be natural to expect an S-matrix upon exchanging momenta; this would be basically built into the two particle Bethe wave function. Hence, to gain some better intuition about the origin of the fundamental relation (\ref{fundamental}) we first need to understand how to move excitations between the different edges of the pentagon. The third and last axiom is precisely about that.
It is depicted in figure \ref{mirror} and reads 
\begin{equation}\label{crossingEq}
P(u^{-\gamma}|v) = \bar P(v| u) 
\end{equation}
where $u^{-\gamma}$ is a mirror transformation such that $E(u^{-\gamma})= -i p(u)$, $p(u^{-\gamma})= -i E(u)$.  
The precise transformation that swaps the energy and momentum depends on which kind of excitation we consider. For the gauge fields under consideration it corresponds to crossing the Zhukowsky cuts, $ x(u^{-\gamma}\pm i/2) = g^2/x(u\pm i/2)$ where $  x(u) = (u+\sqrt{u^2-4 g^2})/2$, see \cite{MoreDispPaper}. Here $g^2=\lambda/(16\pi^2)$ and $\lambda = g_{YM}^2 N_c$ is the 't Hooft coupling.

As a corollary of our axioms, one can easily check  that $\bar P(u^{2\gamma}|v)/P(u^{-3\gamma}|v) = S(v,u)$. This equation has a neat interpretation: we can bring a particle from the bottom to the top of the polygon either through the left by using $u\to u^{2\gamma}$ or through the right through $u\to u^{-3\gamma}$. Both give us two $F$'s on the top but depending on which option we choose we end up with $u$ to the left or to the right of the top excitation $v$. To compare both options we have to permute the two excitations thereby acquiring an $S$-matrix factor. This is an important self-consistency check of our axioms, and provides further motivation for the fundamental relation ({\ref{fundamental}), but it does not provide a derivation of it. At the same time, these kind of manipulations illustrate how we can obtain the transitions from the vacuum to multi-particle states starting from transitions with both top and bottom excitations. For example, according to the discussion above, $\bar P(u^{2\gamma}|v) = P(0|u,v)$. It is quite amusing to see that the single particle transition is related, by analytic continuation, to the two-particle creation amplitude.

We also studied multi-particle transitions. Naturally, they satisfy constraints similar to the three axioms presented above. In addition they should obey bootstrap-like equations that relate their residues to transitions involving a smaller number of particles. As a solution to these equations, we conjecture that the transition of $N$ gauge fields $F$ into $M$ gauge fields $F$ factorizes as
\beq
P({\bf u}|{\bf v}) =\frac{ \prod\limits_{i,j} P(u_i|v_j) }{\prod\limits_{i>j} P(u_i|u_j) \prod\limits_{i<j} P(v_i|v_j)} \,. \label{multi}
\eeq

As mentioned before, the above formulae (\ref{parity}\,-\,\ref{crossingEq})  also apply
 for all other fundamental excitations, up to minor modifications. For example, for scalars, the formulae are even simpler; there is no bar in the r.h.s. of (\ref{crossingEq}). For fermions the r.h.s. of (\ref{fundamental}) should be multiplied by $-1$ while the crossing equation (\ref{crossingEq}) is less well understood. We have conjectures for all these single particle transitions as well as for bound states \cite{toappear}. Below we present in detail the solution for the gluonic transitions. 

\section{Solution for gauge field}
Equations (\ref{parity}), (\ref{fundamental}) and (\ref{crossingEq}) are the fundamental axioms of the Pentagon transitions. We will now present a finite coupling solution to these equations  for the two gluonic transitions $F \to F $ and $F  \to \bar{F} $. The solution reads
\beq
\begin{aligned}\label{voila}
&P(u|v)^2 = \[ \frac{f(u,v)}{g^2 (u-v)(u-v-i)} \]^{\eta} \frac{S(u,v)}{S(u^{\gamma},v)} \, , \\ 
\end{aligned}
\eeq
with $\eta=1$ and the function $f(u, v) = f(v, u)$ given by  
\beqa
f(u, v) = x^{+}x^{-}y^{+}y^{-}\left(1-g^2/x^{+}y^{-}\right)\left(1-g^2/x^{-}y^{+}\right) \nonumber \\ \left(1-g^2/x^{+}y^{+}\right)\left(1-g^2/x^{-}y^{-}\right)\,,\,\,\nonumber
\eeqa
when written in terms of the Zhukowsky variables $x^{\pm} = x(u\pm i/2)$ and $y^{\pm} = x(v\pm i/2)$. For $\bar P(u|v)$ we have the same as in (\ref{voila}) with $\eta =-1$. 
One easily verifies that the expression~(\ref{voila}) solves the relations~(\ref{fundamental}),~(\ref{crossingEq}) using unitarity, $S(u, v)S(v, u) = 1$, the mirror invariance~\cite{AldayMaldacena,OPEpaper} of the flux tube dynamics, $S(u^{\gamma}, v^{\gamma}) = S(u, v)$, and the crossing relation obeyed by the gluon S-matrix, $(u-v-i)S(u^{\gamma}, v) = (u-v+i)S(v^{\gamma}, u)$. The mirror S-matrix $S(u^{\gamma}, v)$ has a simple zero at $u=v$ and therefore $P(u|v)$ has a simple pole as required by the relation to the measure (\ref{measureEq}). 
Equation (\ref{voila}) renders the connection between the Space-time and the Flux-tube S-matrices explicit. 
We now consider the weak and strong coupling limits of these finite coupling conjectures. 

\subsection{Perturbative Regime}
To leading order at weak coupling, 
\beqa\la{perturbative}
P(u|v)&=& -\frac{(u^2+\ft{1}{4})\Gamma(iu-iv)(v^2+\ft{1}{4})}{g^2\,\Gamma(\ft{3}{2}+iu)\Gamma(\ft{3}{2}-iv)} + O(g^0)\, , \\
\bar P(u|v)&=&\frac{\Gamma(2+iu-iv)}{\Gamma(\ft{3}{2}+iu)\Gamma(\ft{3}{2}-iv)} + O(g^2)\, ,\nn
\eeqa
while for the measure \cite{footnoteSign}
\beqa
&&\mu(u) = - \frac{\pi g^2}{(u^2+\ft{1}{4})\cosh{(\pi u)}}\bigg[1+g^2 \bigg(\frac{5\pi^2}{6}+\frac{8u^2-1}{2(u^2+\ft{1}{4})^2}\qquad \\
&&-\frac{1}{2}\(H(\ft{1}{2}+iu)+H(\ft{1}{2}-iu)\)^2-\frac{3\pi^2}{2\cosh^2{(\pi u)}}\bigg)+ \ldots\bigg]\, ,\nn
\eeqa
with $H(z)=\partial_{z}\log{\Gamma(z+1)}+\gamma_E$. 
The fact that $P$ and $\bar P$ start at different loop orders is to be expected since at leading order gluons preserve their helicity, see e.g.  \cite{HexagonOPEpapers}. It is nice to see this feature coming out naturally in this context. 

We shall now describe some perturbative checks of our results. The first one concerns the hexagon Wilson loop. To any order in perturbation theory, its leading OPE behaviour is governed by the exchange of a single twist-one excitation and is given by $\mathcal{W}_\text{hex} = e^{-\tau} f(\tau,\sigma,\phi) + \mathcal{O}(e^{-2\tau})$.
Using our expression (\ref{heptagondecomposition}) we can compute this quantity to all loops. It is essentially governed by the measure and reads
\beq
f(\tau,\sigma,\phi) =2\cos(\phi) \!\int\limits_{-\infty}^{\infty} \frac{du}{2\pi} \,\mu(u)\, e^{-\gamma(u)\tau+i p(u) \sigma}\, , \la{measureHex}
\eeq
where $\gamma(u) = E(u)-1$ is the anomalous energy of the $F$ excitation \cite{BenDispPaper}. Both the measure and the anomalous dimension start at order $g^2$ so that at $l$ loops we get a polynomial of degree $l-1$ in $\tau$. The relation (\ref{measureHex}) suffices to determine the two unfixed constants $\alpha_1$ and $\alpha_2$ in the Hexagon three loop result \cite{Lance} to be exactly as derived by the $\bar Q$ equation in \cite{Qbar}. Similarly, we can compute the leading OPE behaviour for general $n$-sided polygons. For example, for the heptagon $\mathcal{W}_\text{hep}$ we have a double expansion in $e^{-\tau_1}$ and $e^{-\tau_2}$. The term proportional to $e^{-\tau_1-\tau_2}$ is particularly interesting because it is governed by the single gluon transitions. It is given by 
\beqa
&&\!\!\!\!\!\!\!\int \frac{du\,dv}{(2\pi)^2} \,\mu(u)\mu(v)\, e^{-\gamma(u)\tau_1+i p(u) \sigma_1-\gamma(v)\tau_2+i p(v) \sigma_2} \la{hepP} \\
&&  2\[\cos(\phi_1+\phi_2) P(-u|v)+\cos(\phi_1-\phi_2)   \bar P(-u|v) \] \,. \nn
\eeqa
We matched this all loop relation against the symbol of the two loop MHV heptagon \cite{simonHep} and found perfect agreement. We checked the validity of both $P$ and $\bar P$ in perturbation theory. When put together, they provide non-trivial non-perturbative evidence for our ansatz since these two transitions are related by a mirror transformation which is non-perturbative. 

In a plain text file attached to this note, we present explicit perturbative expansions of the energy, momentum, measure $\mu(u)$, and pentagon transition $P(u|v)$, for the gluonic excitation up to order $g^8$. These can be used to predict the leading OPE behaviour of the hexagon and heptagon WLs up to four loops. It is straightforward to expand~(\ref{voila}) further thus producing infinitely many conjectures for an arbitrary high number of loops.

We stress that the relations (\ref{measureHex}),(\ref{hepP}) contain, at any given order in perturbation theory, all the powers of $\tau_i$ multiplying the leading exponential behaviour $e^{-\tau_i}$ and not just the highest power of $\tau_i$. The latter was previously studied in the OPE context and is known as the leading OPE discontinuity. To access the remaining $e^{-n\tau_i}$ contributions, with $n>1$, we need to consider both heavier excitations, such as bound-states of gluons, and multi-particle transitions, such as the ones given by (\ref{multi}). This represents a formidable task, but this is all that is left to compute WL in the planar limit at finite coupling.

\subsection{Strong Coupling}
At strong coupling $\sqrt{\lambda} \gg 1$ the gluonic excitations become relativistic particles of mass $\sqrt{2}$~\cite{FT,AldayMaldacena}. Our expression~(\ref{voila}) predicts that their pentagon transitions simplify drastically in this limit. To leading order, we find that they both become trivial, $P(u|v) \sim \bar P(u|v) \sim 1$, and thus show no dependence on the rapidities. This feature is {essential to match with the string theory prediction \cite{toappear} and is directly related to the exponential form of the amplitude at strong coupling~\cite{AM}.} The subleading correction is also of interest as it turns out to be captured as well by the classical string saddle-point, i.e., by the so called strong coupling Y-system~\cite{AMSV}. It is conveniently written in terms of a kernel $K$ as
\beq\la{srtongcouplingkernel}
P(u|v) = 1+ \frac{2\pi}{\sqrt{\lambda}}K(\theta, \theta') + O(1/\lambda)\, ,
\eeq
whose expression is
\beq
\!\!K(\theta,\theta')\! =\! \frac{i\cosh{(2\theta)}\cosh{(2\theta')}}{2\sinh{(2\theta-2\theta')}}\!\[\sqrt{2}\cosh{(\theta-\theta'-i\ft{\pi}{4})}+\!1\]\, ,
\eeq
and where $\theta$ is the hyperbolic rapidity, related to $u$ by  $u = 2g\tanh{(2\theta)} + O(g^0)$. For $\bar P$ we have
\beq 
\bar K(\theta,\theta')=K(\theta'-i\pi/2,\theta)\,.
\eeq
This follows readily from (\ref{crossingEq}) since at strong coupling the mirror transformation becomes a simple shift of the hyperbolic rapidity by $i\pi/2$.  Similarly for the measure, we find that $\mu(u) = -1+O(1/\sqrt{\lambda})$ and thus
\beq\la{strongcouplingmeasure}
\mu(u)\,\frac{du}{2\pi} =-\frac{\sqrt{\lambda}}{2\pi}\times{d\theta\over\pi\cosh^2(2\theta)}+...\,.
\eeq

Let us now demonstrate how these expressions are reproduced from the strong coupling Y-system solution. As found in \cite{AMApril,AMSV,OPEpaper}, 
the strong coupling result reads
\beq
\log\<W\>=-{\sqrt\lambda\over2\pi}\[A_\text{div}+A_\text{BDS like}+A_0+YY_{cr}\]
\eeq
where $A_\text{div}$, $A_\text{BDS like}$ and $A_0$ are simple, explicitly known, functions of both the positions of the cusps of the polygon and the UV cut-off. The most nontrivial part of the result is the critical Yang-Yang functional $YY_{cr}$. Remarkably, when considering our ratio defined in figure \ref{calW}, it is this part and this part only that remains,
\beq {\cal W}=\exp\[-{\sqrt\lambda\over2\pi} YY_{cr}\]\,,
\eeq
while all the other contributions cancel out exactly.
Now, to read off the transitions and measures we expand $\cal W$ at large $\tau_i$ using the Y-system prediction for $YY_{cr}$. It turns out that there is no order of limits issue:
the various kernels $K_{ab}(\theta,\theta')$ and measures $\mu_a(\theta)$ obtained from our expressions as in (\ref{strongcouplingmeasure}) and (\ref{srtongcouplingkernel}) are in direct correspondence with the measures and (linear combination of the) kernels appearing in the strong coupling Thermodynamic Bethe Ansatz (TBA), see appendix F of \cite{OPEpaper}. For example, the gluonic measure (\ref{strongcouplingmeasure}) governs the leading large $\tau$ contribution {(of the mass $\sqrt 2$ excitation)} to the hexagon WL and it matches perfectly with the stringy result $R_{\sqrt 2}$ given in equation (4.2) in \cite{OPEpaper}. 

{We should stress nevertheless} that the derivation of the strong coupling $YY_{cr}$ from the decomposition (\ref{decomposition})  {is far from being trivial}. For example, poles in the transitions pinch in the contours of integration at large $\lambda$, obliging us to rearrange slightly (\ref{decomposition}). It illustrates nicely how important it is to start with finite coupling. 

Finally,  it is worth emphasizing that the strong coupling result in \cite{AMSV} is a solution of a minimal area problem \cite{AM}. Its original derivation was purely geometrical, with no direct mention of the OPE. Still, the strong coupling result -- and its associated TBA equations -- was begging for such a description \cite{AGM,OPEpaper}. It is rewarding to finally have an alternative derivation that puts weak and strong coupling on exactly the same footing. More details will be presented in \cite{toappear}.

\vspace{-2mm}

\section{Discussion}

\vspace{-2mm}

The decomposition (\ref{decomposition}) breaks down the computation of Scattering Amplitudes in planar $\mathcal{N}=4$ SYM into fundamental building blocks which we dubbed \textit{Pentagon transitions}. These transitions obey a set of bootstrap equations (\ref{parity}\,-\,\ref{crossingEq}) that can be solved {thanks to} the Integrability of the GKP flux tube.

In this note we presented a conjecture for all gluonic transitions at any value of the coupling, see (\ref{multi}) and (\ref{voila}). We have similar conjectures for all the single particle transitions as well as for bound states \cite{toappear}. The derivation of the single scalar transitions is technically simpler than what we explained in this paper while the fermion transition appears more complicated due to a lack of understanding of its crossing transformation. 
For multi-particle transitions involving scalars and fermions
 there is, in addition to a (known) dynamical factor as in (\ref{multi}), an R-symmetry matrix structure that can hopefully be fixed using Integrability. 
 For MHV amplitudes, the gluonic transitions considered in this paper dominate both at weak and strong coupling. For N$^k$MHV amplitudes, scalar and/or fermion transitions will certainly play a more important role \cite{superloop,toappear}. The OPE approach exposes the integrability of the problem but not all symmetries are manifest since some are broken by the GKP flux tube. 
It would be interesting to draw inspiration from other approaches \cite{YangianOfDrummondHennPlefka,Grass} that make these symmetries manifest. 

Once all transitions have been found, the next obvious question is whether the decomposition (\ref{decomposition}) can be re-summed into something akin to
a Thermodynamic Bethe Ansatz partition function. Encouraging evidence comes from strong coupling where this is possible. We can now study sub-leading corrections at strong coupling in a controllable way and see what happens to this re-summation. This could provide important hints regarding the full finite coupling structure. 
Of course, to evaluate the amplitude at finite coupling on a computer, it is not crucial whether the decomposition (\ref{decomposition}) admits a nice analytical resummation or not. 

Finally, it would be very interesting to study what happens in other theories or beyond the planar limit.  \\

{\it Acknowledgements:} We 
thank Y.~Fan for collaboration at initial stages of this project. We thank F.~Cachazo, J.~Caetano, L.~Dixon, D.~Gaiotto, J.~Henn, G.~ Korchemsky, J.~Maldacena, J.~Toledo and T.~Wang for discussions. We are very grateful to J.~Bourjaily, S.~Caron-Huot and S.~He for discussions and for providing us with invaluable non-MHV data used to check several conjectures. Research at the Perimeter Institute is supported in part by the Government of Canada through NSERC and by the Province of Ontario through MRI. The research of A.S. has been supported in part by the Province of Ontario through ERA grant ER 06-02-293 and by the U.S. Department of Energy grant DE-FG02-90ER40542.



\begin{thebibliography}{99}
\bibitem{footnoteSign}
The measure as currently defined through (\ref{measureEq}) is not positive definite for all excitations. We could make it positive at the expense of redefining the angles $\phi_i$ and introducing some slightly unwelcome minus signs in (\ref{measureEq}).
\end{thebibliography}
\end{document}